\documentstyle[aps,prb,preprint]{revtex}

\begin{document}
\author{{\bf Miodrag} {\bf L.} {\bf Kuli\'{c}}$^{1}${\bf \ and Oleg V. Dolgov}$^{2}$}
\address{$^{1}$Physikalisches Institut, Universit\"{a}t Bayreuth, 95440 Bayreuth,\\
Germany\\
$^{2}$Institut f\"{u}r Theoretische Physik, Universit\"{a}t T\"{u}bingen,\\
Germany}
\title{Anisotropic impurities in anisotropic superconductors}
\date{today}
\maketitle

\begin{abstract}
Physical properties of anisotropic superconductors like the critical
temperature and others depend sensitively on the electron mean free path.
The sensitivity to impurity scattering and the resulting anomalies are
considered a characteristic feature of strongly anisotropic pairing. These
anomalies are usually analyzed in terms of s-wave impurity scattering which
leads to universal pair breaking effects depending on only two scattering
parameters, the mean free path and the impurity cross section. We
investigate here corrections coming from anisotropies in the scattering
cross section, and find not only quantitative but also qualitative
deviations from universal s-wave isotropic pairbreaking. The properties we
study are the transition temperature, the density of states, quasiparticle
bound states at impurities, and pinning of flux lines by impurities.\newline
\end{abstract}


\newpage

\section{\ Introduction}

Unconventional anisotropic pairing is evidently realized in high temperature
superconductors ($HTS$) - see review \cite{KulicRev}, and probably in some
heavy Fermion superconductors ($HFS$) - see review \cite{Sauls}. There are
good evidences for d-wave pairing in optimally doped $HTS$ oxides \cite
{KulicRev} but the type of pairing in HFS is still unclear \cite{Sauls}. The
effect of impurities on unconventional pairing is an important tool in
analyzing the symmetry of the pairing amplitude, and is the subject of a
number of experimental and theoretical works \cite{Pokrovskii}, \cite{Hotta}%
. Most of calculations were done assuming an s-wave impurity scattering
potential $u_{imp}({\bf p},{\bf p}^{\prime })=const$, and taking either the
Born limit ($N(0)u_{imp}\ll 1$) or unitarity limit ($N(0)u_{imp}\gg 1$).

Surprisingly, a number of experiments on the optimally $HTS$ oxides have
shown that $d-wave$ pairing is quite robust, i.e. not very sensitive to
various kinds of impurities and defects. For instance, the decrease of the
critical temperature $T_{c}(\rho _{imp})$, with increasing residual
resistivity, $\rho _{imp}$, is much smaller than the theory with the $s-wave$
impurity scattering predicts \cite{Pokrovskii}, \cite{Hotta}, \cite{Sun1}. A
way out of this experimental and theoretical discrepancy of pair-breaking
effects by impurities in $HTS$ oxides was proposed by the authors of Refs. 
\cite{KuOudo}, \cite{Haran}, who invoked a momentum dependent impurity
scattering potential with an appreciable contribution in the d-channel. The
microscopic theory in Ref. \cite{KuOudo} accounts for the renormalization of
the impurity potential by strong correlations, which gives rise to a
pronounced forward scattering peak, while backward scattering is suppressed,
as first proposed in Ref. \cite{Kulic1} (see also \cite{KulicRev}).
Application of this theory to impurity scattering \cite{Kulic1} shows that
in addition to the contribution in the s-channel there is a significant
contribution to the Born amplitude from the d-channel of the same magnitude
, in particular for low (hole) doping concentration, $\delta <0.2$. As a
consequence, the decrease of $T_{c}(\rho _{imp})$ with increasing $\rho
_{imp}$ is much slower than the theory with exclusively s-wave impurity
scattering predicts \cite{Hotta}. This renormalization effect explains the
robustness of d-wave pairing in $HTS$ oxides.

One may rise the question whether this robustness also holds far away from $%
T_{c}$ and for very strong scattering potential, for instance in the
unitarity limit. To answer this question we shall analyze a class of models
by calculating the scattering $T$-matrix with an impurity potential that
depends on the scattering angles.

A related class of problems, which we study in section III, is related to
the impurity scattering in the two-band model. Recently, several models for
the pairing mechanism in HTS oxides based on two-band and multi-band models 
\cite{Hofmann}, \cite{Gajic}, \cite{Golubov}, \cite{Combescot} were
suggested, and impurity effects studied in Born approximation. Magnetic and
non-magnetic interband scattering can lead in this model to a lowering of
the critical temperature and also to a relative sign change of the order
parameters in different bands \cite{Golubov}. In section III we analyze the
changes in the two-band model when going beyond the Born limit. It was shown
that in the unitarity limit the Anderson theorem holds.

In previous sections we studied a homogeneous superconductor with
homogeneously distributed impurities. Selected inhomogeneous problems are
studied in section IV, such as the bound states at an impurity, and the
pinning energy at an impurity (defect) of singly- and double-quantized
vortices.

\section{Anisotropic scattering in anisotropic and homogeneous
superconductors}

In the following we analyze superconducting properties of anisotropic
superconductors in the presence of momentum-dependent nonmagnetic impurity
scattering by the quasiclassical equations of Eilenberger,
Larkin-Ovchinnikov \cite{Eilenberger}, \cite{Larkin} ($ELO$ equations ). For
a homogeneous distribution the quasiclassical Green's function matrix, $\hat{%
g}({\bf p}_{F},{\bf R},\omega _{n})$, is independent of ${\bf R}$, and the
quasiclassical equations read 
\begin{equation}
\lbrack i\omega _{n}\hat{\tau}_{3}-\hat{\Delta}({\bf p}_{F},\omega _{n})-%
\hat{\sigma}_{imp}({\bf p}_{F},\omega _{n}),\hat{g}({\bf p}_{F},\omega
_{n})]=0  \label{elohom}
\end{equation}
\begin{equation}
\hat{g}^{2}({\bf p}_{F},\omega _{n})=-\hat{1}.  \label{norm}
\end{equation}

We assume weak-coupling superconductivity with $\hat{\Delta}({\bf p}%
_{F})(=i\Delta ({\bf p}_{F})\hat{\tau}_{2})$, where $\Delta ({\bf p}_{F})$
is real. The $2\times 2$ matrices$\ \hat{\tau}_{0}\equiv \hat{1}$ and $\hat{%
\tau}_{1,2,3}$ are Nambu-Gor'kov matrices. The effect of nonmagnetic
impurities is described by the self-energy, $\hat{\sigma}_{imp}$, given in
terms of the forward scattering part, $\hat{t}({\bf p}_{F},{\bf p}%
_{F}^{\prime },\omega _{n})$, of the T-matrix \cite{Eilenberger}, \cite
{Larkin}, \cite{Serene} 
\begin{equation}
\hat{\sigma}_{imp}({\bf p}_{F},\omega _{n})=c\hat{t}({\bf p}_{F},{\bf p}%
_{F},\omega _{n}),  \label{sigma}
\end{equation}
where $c(\ll 1)$ is the impurity concentration.

For simplicity we assume an isotropic Fermi surface but pairing and impurity
scattering are angle-dependent, i.e. $\Delta ({\bf p}_{F})\equiv \Delta (%
{\bf s})$, $\hat{g}({\bf p}_{F},\omega _{n})\equiv \hat{g}({\bf s},n)$ and $%
\hat{t}({\bf p}_{F},{\bf p}_{F}^{\prime },\omega _{n})\equiv \hat{t}({\bf s,s%
}^{\prime }{\bf ,}n)$ where ${\bf s=p}_{F}/p_{F}$.The T-matrix is the
solution of the equation 
\begin{equation}
\hat{t}({\bf s,s}^{\prime }{\bf ,}n)=u({\bf s,s}^{\prime })\hat{1}+N(0)\int d%
{\bf s}^{\prime \prime }u({\bf s,s}^{\prime \prime })\hat{g}({\bf s}^{\prime
\prime },n)\hat{t}({\bf s}^{\prime \prime }{\bf ,s}^{\prime }{\bf ,}n),
\label{tmatrix}
\end{equation}
where for the $2D$ systems, which we consider here, one has $\int d{\bf %
s\{..\}}\equiv \int {\bf \{..\}}d\theta /2\pi $. Since $\Delta ({\bf s})$ is
real one has $\hat{g}=g_{2}\hat{\tau}_{2}+g_{3}\hat{\tau}_{3}$ and $\hat{t}$
is given by $\hat{t}=t_{0}\hat{\tau}_{0}+t_{1}\hat{\tau}_{1}+t_{2}\hat{\tau}%
_{2}+t_{3}\hat{\tau}_{3}$.

Because the unperturbed solution has the form ($\omega _{n}=\pi T(2n+1)$) 
\begin{equation}
\hat{g}^{(0)}({\bf s},n)=-\frac{i\omega _{n}\hat{\tau}_{3}-i\Delta _{0}({\bf %
s})\hat{\tau}_{2}}{\sqrt{\omega _{n}^{2}+\Delta _{0}^{2}({\bf s})}}
\label{g0}
\end{equation}
then $\hat{g}({\bf s},n)$ is searched for in the form 
\begin{equation}
\hat{g}({\bf s},n)=-\frac{i\tilde{\omega}_{n}({\bf s})\hat{\tau}_{3}-i\tilde{%
\Delta}({\bf s,}\omega _{n})\hat{\tau}_{2}}{\sqrt{\tilde{\omega}_{n}^{2}(%
{\bf s})+\tilde{\Delta}^{2}({\bf s,}\omega _{n})}},  \label{g}
\end{equation}
where 
\begin{equation}
\tilde{\omega}_{n}({\bf s})=\omega _{n}({\bf s})+ic_{i}t_{3}({\bf s,s,}n)
\label{omega}
\end{equation}
\begin{equation}
\tilde{\Delta}({\bf s,}\omega _{n})=\Delta ({\bf s})-ic_{i}t_{2}({\bf s,s,}%
n).  \label{delta}
\end{equation}
The self-consistency equation for $\Delta ({\bf s})$ is given by 
\begin{equation}
\Delta ({\bf s})=N(0)T\sum_{n}\int d{\bf s}^{\prime }V{\bf (s,s}^{\prime }%
{\bf )}g_{2}({\bf s}^{\prime },n),  \label{selfcons}
\end{equation}
where the pairing potential $V_{p}{\bf (s,s}^{\prime }{\bf )=}V_{p}Y({\bf s}%
)Y({\bf s}^{\prime })$ is assumed in the factorized form with $<Y^{2}({\bf s}%
)>_{{\bf s}}=1$. The latter implies that the order parameter has the form $%
\Delta ({\bf s})=\Delta \cdot Y({\bf s})$. For convenience we define $\Gamma
_{u}(\equiv c\gamma _{u})=c(\pi N(0))^{-1}$ and $v({\bf s,s}^{\prime
})\equiv \pi N(0)u({\bf s,s}^{\prime })$. In what follows we consider the
effects of anisotropic impurity scattering on the anisotropic pairing where $%
<Y({\bf s})>_{{\bf s}}=0$.

{\bf 1}. {\bf Anisotropic impurity scattering and nodeless anisotropic
pairing}

First, we consider the nodeless d-wave like pairing $\Delta ({\bf s})=\Delta
\cdot Y({\bf s})$ which is characterized by $\langle Y({\bf s})\rangle _{%
{\bf s}}=0$ and $Y^{2}({\bf s})=1$. This means that there is a finite gap
everywhere on the Fermi surface, i.e. $\Delta ({\bf s})\neq 0$. It is
interesting to mention, that besides the simplicity of this kind of pairing
and its adequacy in some qualitative understanding of d-wave pairing it also
appears to be a solution of the spin-bag model \cite{SWZ} for $HTS$ oxides.
In this model the nodeless $d-wave$ like pairing is due to residual
(longitudinal and transverse) spin fluctuations on the antiferromagnetic
background, where the $AF$ order is distorted locally by hole doping and the
spin-bag is formed around doped holes. The impurity scattering potential is
assumed to have the form 
\begin{equation}
v({\bf s,s}^{\prime })=v_{0}+v_{2}Y({\bf s})Y({\bf s}^{\prime }),
\label{vsep}
\end{equation}
i.e. it contains an anisotropic contribution in the same channel as the
unconventional pairing. The solution of $Eq.(\ref{tmatrix})$ for $t_{3}$ and 
$t_{2}$ is searched in the form 
\begin{equation}
t_{3}({\bf s,s}^{\prime })=[\tilde{t}_{30}(n)+\tilde{t}_{32}(n)Y({\bf s})Y(%
{\bf s}^{\prime })]g_{3},  \label{t3}
\end{equation}
\begin{equation}
t_{2}({\bf s,s}^{\prime })=\tilde{t}_{2}(n)[g_{2}({\bf s},n)+g_{2}({\bf s}%
^{\prime },n)].  \label{t2}
\end{equation}
(Note, in this model $g_{2}({\bf s},n)=\tilde{g}_{2}(n)Y({\bf s})$, $%
g_{2}^{2}({\bf s},n)=\tilde{g}_{2}^{2}(n)$, $g_{3}({\bf s},n)=g_{3}(n)$ and
due to $Eq.(\ref{norm})$ one has $g_{3}^{2}({\bf s},n)+\tilde{g}_{2}^{2}(%
{\bf s},n)=-1$.) The solution is given by 
\begin{equation}
\tilde{t}_{30}(n)=\gamma _{u}v_{0}^{2}\frac{1+v_{2}^{2}}{%
(1+v_{0}^{2})(1+v_{2}^{2})+(v_{0}-v_{2})^{2}\tilde{g}_{2}^{2}(n)},
\label{t30}
\end{equation}
\begin{equation}
\tilde{t}_{2}(n)=\gamma _{u}v_{0}v_{2}\frac{1+v_{0}v_{2}}{%
(1+v_{0}^{2})(1+v_{2}^{2})+(v_{0}-v_{2})^{2}\tilde{g}_{2}^{2}(n)},
\label{t2n}
\end{equation}
while $\tilde{t}_{32}(n)=\tilde{t}_{30}(n,v_{0}\leftrightarrow v_{2})$. \
Several interesting results comes out in this case.

({\bf a}) {\bf The} {\bf critical temperature} $T_{c}$

In the limit $T\rightarrow T_{c}$ $Eqs.(\ref{selfcons},\ref{t30}-\ref{t2n})$
become 
\begin{equation}
\ln \frac{T_{c}}{T_{c0}}=\Psi (\frac{1}{2})-\Psi (\frac{1}{2}+\frac{\Gamma
_{pb}}{2\pi T_{c}}),  \label{tc}
\end{equation}
where the pair-breaking parameter $\Gamma _{pb}$ is given by ($\Gamma
_{u}=c\pi /N(0)$) 
\begin{equation}
\Gamma _{pb}=\Gamma _{u}\frac{(v_{0}-v_{2})^{2}}{(1+v_{0}^{2})(1+v_{2}^{2})}.
\label{gamapb}
\end{equation}
Note that $T_{c}$ vanishes for $\Gamma _{pb}^{c}\approx 0.88T_{c0}$ and this
pairing is in some respects similar to d-wave pairing. It is apparent from $%
Eqs.(\ref{tc}-\ref{gamapb})$ that the pair-breaking effect of impurities is
weakened in the presence of momentum-dependent scattering and it is even
zero for $v_{0}=v_{2}$. The latter result has been previously derived in the
Born approximation \cite{KuOudo}. For $v_{2}\approx v_{0}$ the slope $%
dT_{c}/d\rho _{imp}$ can be very small even for appreciable values of $\rho
_{imp}\sim \Gamma _{tr}=\Gamma _{u}(\bar{\sigma}_{0}+\bar{\sigma}_{2}),$
because in that case $\Gamma _{pb}\ll \Gamma _{tr}$ as indicate the
experimental results of \cite{Sun1}. The parameters $\bar{\sigma}_{i}$ are
given by 
\begin{equation}
\bar{\sigma}_{i}=\frac{v_{i}^{2}}{1+v_{i}^{2}}\text{, \ \ }i=0,1,2...
\label{sig}
\end{equation}

The resistivity, $\rho _{imp}$, and the reduction of $T_{c}$ due to impurity
scattering, $T_{c}(\rho _{imp})$, depend on the classical transition rate, $%
W({\bf s,s}^{\prime })=\Gamma _{u}\mid t_{N}({\bf s,s}^{\prime },n)\mid ^{2}$%
in the normal state \cite{Rainer}. This transition rate comprises all the
needed information on impurity scattering for either solving the normal
state Boltzmann equation to determine $\rho _{imp}$ or the linearized gap
equation $Eq.(\ref{selfcons})$ to determine $T_{c}$. For the latter purpose
one needs linear (integral) equation for $g_{2}({\bf s},n)$ which reads

\begin{equation}
\mid \omega _{n}\mid g_{2}({\bf s},n)-\Delta ({\bf s})+\int d{\bf s}^{\prime
}W({\bf s,s}^{\prime })[g_{2}({\bf s},n)-g_{2}({\bf s}^{\prime },n)]=0,
\label{g2w}
\end{equation}
where the normal state t-matrix $t_{N}({\bf s,s}^{\prime },n)$ is the
solution of the equation 
\begin{equation}
t_{N}({\bf s,s}^{\prime },n)=v({\bf s,s}^{\prime })-i\pi sign(\omega
_{n})\int d{\bf s}^{\prime \prime }v({\bf s,s}^{\prime \prime })t_{N}({\bf s}%
^{\prime \prime }{\bf ,s}^{\prime },n).  \label{tN}
\end{equation}
Hence, measurements of $T_{c}(\rho _{imp})$ curve carry not enough
informations on the microscopic scattering data, i.e. the scattering
T-matrix. More such informations are contained in spectroscopic data on
anisotropic superconductors at temperatures $T\ll T_{c}$, such as tunneling
data or optical data at about gap frequency.

{\bf (b) }The density of states{\bf , }$N(\omega )=N(0)%
\mathop{\rm Im}%
\int d{\bf s}$ $g_{3}({\bf s,}i\omega _{n}\rightarrow \omega -i\eta )$,
depends in the presence of pair-breaking impurities significantly on the
values $v_{0}$ and $v_{1}$. It is known \cite{Preosti} that in the case of
s-wave scattering only ($v_{2}=0$) one has $N(\omega =0)\neq 0$ for $\Gamma
_{u}v_{0}^{2}>\Delta $, and the highest value, $N(\omega
=0)=N(0)/[0.5+0.5(1+(2\Delta /\Gamma )^{2})^{1/2}]^{1/2}$, where $\Gamma
\equiv \Gamma _{u}\bar{\sigma}_{0}$, is reached in the unitarity limit. On
the other hand, one obtains in the limiting case, $v_{0}=v_{2}$, a
restoration of the gap, $N(\omega =0)=0$. Despite of the strong scattering
limit $N(\omega )$ is BCS-like.

{\bf 2. Isotropic impurity scattering and }${\bf d-wave}${\bf \ pairing}

Let us study a two-dimensional superconductor with the pairing function $%
\Delta ({\bf s})\equiv \Delta (\varphi )(=\Delta \cdot Y_{2}(\varphi ))\sim
\cos 2\varphi $ - $d-wave$ pairing. Note this case seems to be more
realistic for HTS oxides than the previous one, because the ''$\cos 2\varphi 
$'' pairing has nodes at the simply connected Fermi surface. We assume that
the isotropic impurity potential depends on the transferred scattering angle 
\begin{equation}
v(\varphi ,\varphi ^{\prime })=v_{0}+2v_{1}\cos (\varphi -\varphi ^{\prime
})+2v_{2}\cos 2(\varphi -\varphi ^{\prime }),  \label{vfi}
\end{equation}
where $v(\varphi ,\varphi ^{\prime })$ contains the pairing channel ($\sim
Y_{2}(\varphi )Y_{2}(\varphi ^{\prime })$) too. (This problem but with $%
v_{2}=0$ is studied in \cite{Haas} but there is an inappropriate sign in the 
$t_{2}$-matrix, which in fact corresponds to a magnetic impurity
scattering). From $Eqs.(\ref{t30}-\ref{t2n})$ one obtains the pair-breaking
parameter $\Gamma _{pb}$

\begin{equation}
\Gamma _{pb}=\Gamma _{u}[\bar{\sigma}_{0}\frac{(1-\alpha )^{2}+\alpha
^{2}(1+v_{0}^{2})}{1+\alpha ^{2}v_{0}^{2}}+\bar{\sigma}_{1}],  \label{gamafi}
\end{equation}
where $\alpha =v_{2}/v_{0}$ and $\bar{\sigma}_{i}$ are given by $Eq.(\ref
{sig})$.

In order to analyze $T_{c}(\rho _{imp})$ dependence, where the residual
impurity resistivity $\rho _{imp}\sim \Gamma _{tr}$, we need the transport
scattering $\Gamma _{tr}$ which is in this case given by 
\begin{equation}
\Gamma _{tr}=\Gamma _{u}\{\bar{\sigma}_{0}^{2}+2\bar{\sigma}_{1}^{2}+2\bar{%
\sigma}_{2}^{2}-2\bar{\sigma}_{0}[\bar{\sigma}_{1}(1+\frac{1}{v_{0}v_{1}})+%
\bar{\sigma}_{2}(1+\frac{1}{v_{0}v_{2}})]\}\text{.}  \label{gamatrans}
\end{equation}

If one wants to interpret depairing effects of impurities and robustness of
pairing in HTS oxides in terms of the above results then the experiments 
\cite{Sun1} imply that the ratio, $\Gamma _{pb}/\Gamma _{tr}$, should be
minimum. In the case when $v_{2}\ll v_{1}$ one obtains, $\Gamma _{pb}/\Gamma
_{tr}=2$, in both, the Born and unitarity, $v_{0},v_{1}\rightarrow \infty $,
limits. For, $v_{1}\ll v_{2},$ $\ $the pair-breaking parameter, $\Gamma
_{pb} $, is minimized for $\alpha =1/2$ which gives, $\Gamma _{pb}/\Gamma
_{tr}\approx 1/3$, in both limits. This means that the latter case is more
appropriate candidate, than the case, $v_{2}\ll v_{1}$, for the qualitative
explanation of robustness of d-wave pairing in $HTS$ oxides.

\section{\bf Two-band model with nonmagnetic impurities}

The interest in two(multi)-band models and in the impurity effects is
renewed after the discovery of $HTS$ oxides \cite{Hofmann}, \cite{Gajic}, 
\cite{Combescot}, \cite{Golubov}, where various kinds of the intra- and
inter-band pairing and impurity scattering are considered. By assuming only
intra-band pairing $\Delta _{\alpha }$ ($\alpha =1,2$) the effect of the
nonmagnetic impurities in the Born approximation is described by the
equations ($n$ enumerates Matsubara frequencies) 
\begin{equation}
\tilde{\omega}_{\alpha n}=\omega _{n}+\sum_{\beta }\frac{\tilde{\omega}%
_{\beta n}}{2Q_{\beta n}}\gamma _{\alpha \beta }  \label{omega2b}
\end{equation}
\begin{equation}
\tilde{\Delta}_{\alpha n}=\Delta _{\alpha }+\sum_{\beta }\frac{\tilde{\Delta}%
_{\beta n}}{2Q_{\beta n}}\gamma _{\alpha \beta }  \label{delta2b}
\end{equation}
\begin{equation}
\Delta _{\alpha }=\pi T\sum_{\beta ,n}^{-\omega _{D}<\omega _{n}<\omega
_{D}}\lambda _{\alpha \beta }\frac{\tilde{\Delta}_{\beta n}}{Q_{\beta n}},
\label{self2b}
\end{equation}
where $Q_{\alpha n}=\sqrt{\tilde{\omega}_{\alpha n}^{2}+\tilde{\Delta}%
_{\alpha n}^{2}}$, $\gamma _{\alpha \beta }=u_{\alpha \beta }^{2}N_{\beta
}(0)$ for the nonmagnetic impurity scattering and $\lambda _{\alpha \beta
}=V_{\alpha \beta }^{p}N_{\beta }(0)$ are corresponding coupling constants.
In Ref. \cite{Golubov} are considered various possibilities for the
suppression of the critical temperature, as well as the relative sign of $%
\Delta _{1}$ and $\Delta _{2}$, in the Born limit for nonzero values of $%
\lambda _{\alpha \beta }$ and $\gamma _{\alpha \beta }$. We note some
interesting conclusions obtained in Born approximation which shall be
compared with the results obtained in the unitarity limit: $(i)$ the
diagonal scattering rate $\gamma _{11}$ and $\gamma _{22}$ disappear from
the linearized $Eq.(\ref{self2b})$ for $T_{c}$; $(ii)$ in the case $\lambda
_{11}\neq 0,\lambda _{22}=\lambda _{12}=\lambda _{21}=0$ the depression of $%
T_{c}$ is given by $\delta T_{c}/T_{c}=-\pi \gamma _{12}/8T_{c}$; $(ii)$ for 
$\lambda _{11}=\lambda _{22}\neq 0$ and $\lambda _{12}=\lambda _{21}=\lambda
_{\perp }<0$ one has $sign(\Delta _{1}/\Delta _{2})=-1$ and $\delta
T_{c}/T_{c}=-\pi (\gamma _{12}+\gamma _{21})/8T_{c}$, while the sign of $%
\Delta _{1}$ and $\Delta _{2}$ is unchanged by the impurities.

The $t$-matrix equation in the two-band model has the form ($\alpha ,\beta
,\gamma =1,2$) ( we consider a rather small impurity concentration and
neglect an interband hybridization)

\begin{equation}
{\bf \hat{t}}(n)={\bf \hat{u}}+\sum_{\gamma }{\bf \hat{u}N}(0){\bf \hat{g}}%
(n){\bf \hat{t}}(n),  \label{t2b}
\end{equation}
where ${\bf \hat{t}}(n)=\sum_{i=0}^{3}{\bf t}_{i}{\bf \otimes }\hat{\tau}%
_{i} $, ${\bf \hat{g}}(n)={\bf g}_{3}{\bf \otimes }\hat{\tau}_{3}+{\bf g}_{2}%
{\bf \otimes }\hat{\tau}_{2}$ and ${\bf \otimes }$ is the direct product of
matrices in the band space (bold) and in the Nambu space (hat). ${\bf g}_{2},%
{\bf g}_{3}$ and ${\bf N}(0)$ are diagonal matrices in the band space.

In the case of nonmagnetic impurities one has ${\bf \hat{u}}^{N}={\bf u}^{N}%
{\bf \otimes }\hat{\tau}_{0}$ and since ${\bf g}_{1}=0$ one has 
\[
{\bf t}_{0}^{N}(n)={\bf u}^{N}+{\bf u}^{N}{\bf N}(0)[{\bf g}_{3}(n){\bf t}%
_{3}^{N}(n)+{\bf g}_{3}(n){\bf t}_{3}^{N}(n)] 
\]
\[
{\bf t}_{1}^{N}(n)={\bf u}^{N}{\bf N}(0)[-i{\bf g}_{3}(n){\bf t}_{2}^{N}(n)+i%
{\bf g}_{2}(n){\bf t}_{3}^{N}(n)] 
\]
\[
{\bf t}_{2}^{N}(n)={\bf u}^{N}{\bf N}(0)[i{\bf g}_{3}(n){\bf t}_{1}^{N}(n)+%
{\bf g}_{2}(n){\bf t}_{0}^{N}(n)] 
\]
\begin{equation}
{\bf t}_{3}^{N}(n)={\bf u}^{N}{\bf N}(0)[{\bf g}_{3}(n){\bf t}_{0}^{N}(n)-i%
{\bf g}_{2}(n){\bf t}_{1}^{N}(n)]  \label{ti2b}
\end{equation}
Let us consider for simplicity the case when $u_{11}^{N},u_{22}^{N}=0$ but
interband scattering, $u_{12}^{N}=u_{21}^{N}=u$ $\neq 0$, and introduce
three parameters 
\begin{equation}
\sigma =\frac{\pi ^{2}N_{1}(0)N_{2}(0)u^{2}}{1+\pi ^{2}N_{1}(0)N_{2}(0)u^{2}}
\label{sig2b}
\end{equation}
and 
\begin{equation}
\Gamma _{i}=\frac{c\cdot \sigma }{\pi N_{i}(0)},\text{ \ }i=1,2.
\label{Gama2b}
\end{equation}
After some straightforward calculations one obtains the renormalized
frequencies, $\tilde{\omega}_{in}$, and order parameters $\tilde{\Delta}%
_{in} $ 
\begin{equation}
\tilde{\omega}_{1n}=\omega _{n}+\Gamma _{1}\frac{(\sigma -1)(\tilde{\omega}%
_{1n}^{2}+\tilde{\Delta}_{1n}^{2})\tilde{\omega}_{2n}-\sigma \tilde{\omega}%
_{1n}\sqrt{\tilde{\omega}_{1n}^{2}+\tilde{\Delta}_{1n}^{2}}\sqrt{\tilde{%
\omega}_{2n}^{2}+\tilde{\Delta}_{2n}^{2}}}{\det 1}  \label{omega1}
\end{equation}
\begin{equation}
\tilde{\Delta}_{1n}=\Delta _{1}+\Gamma _{1}\frac{(\sigma -1)(\tilde{\omega}%
_{1n}^{2}+\tilde{\Delta}_{1n}^{2})\tilde{\Delta}_{2n}-\sigma \tilde{\Delta}%
_{1n}\sqrt{\tilde{\omega}_{1n}^{2}+\tilde{\Delta}_{1n}^{2}}\sqrt{\tilde{%
\omega}_{2n}^{2}+\tilde{\Delta}_{2n}^{2}}}{\det 1},  \label{delta1}
\end{equation}
where 
\[
\det 1=2(\sigma -1)\sigma \sqrt{\tilde{\omega}_{1n}^{2}+\tilde{\Delta}%
_{1n}^{2}}(\tilde{\Delta}_{1n}\tilde{\Delta}_{2n}+\tilde{\omega}_{1n}\tilde{%
\omega}_{2n})- 
\]
\begin{equation}
-[2(\sigma -1)\sigma +1](\tilde{\omega}_{1n}^{2}+\tilde{\Delta}_{1n}^{2})%
\sqrt{\tilde{\omega}_{2n}^{2}+\tilde{\Delta}_{2n}^{2}}.  \label{det1}
\end{equation}
The solution for the second band is obtained from $Eqs.(\ref{omega1}-\ref
{det1})$ by replacing $1\Longleftrightarrow 2$. In the Born limit one gets 
\begin{equation}
\tilde{\omega}_{1n}=\omega _{n}+\Gamma _{1}\frac{\sigma \tilde{\omega}_{2n}}{%
\sqrt{\tilde{\omega}_{2n}^{2}+\tilde{\Delta}_{2n}^{2}}}  \label{omegab}
\end{equation}
\begin{equation}
\tilde{\Delta}_{1n}=\Delta _{1}+\Gamma _{1}\frac{\sigma \tilde{\Delta}_{2n}}{%
\sqrt{\tilde{\omega}_{2n}^{2}+\tilde{\Delta}_{2n}^{2}}},  \label{deltab}
\end{equation}
i.e. the interband scattering mixes both bands. In the unitarity limit $%
\sigma \rightarrow 1$ ($u\rightarrow \infty $) the bands are decoupled, i.e. 
\begin{equation}
\tilde{\omega}_{\alpha n}=\omega _{n}+\Gamma _{\alpha }\frac{\tilde{\omega}%
_{\alpha n}}{\sqrt{\tilde{\omega}_{\alpha n}^{2}+\tilde{\Delta}_{\alpha
n}^{2}}}  \label{omegauni}
\end{equation}
\begin{equation}
\tilde{\Delta}_{\alpha n}=\Delta _{\alpha n}+\Gamma _{\alpha }\frac{\tilde{%
\Delta}_{\alpha n}}{\sqrt{\tilde{\omega}_{\alpha n}^{2}+\tilde{\Delta}%
_{\alpha n}^{2}}}.  \label{deltauni}
\end{equation}
So, in this case the Anderson theorem is restored, i,e, the thermodynamic
properties are impurity independent.

The latter result can be generalize to the case 
\begin{equation}
{\bf u}^{N}=\left( 
\begin{array}{cc}
\alpha u & u \\ 
u & u_{22}
\end{array}
\right) .  \label{un}
\end{equation}
For $u\rightarrow \infty $ but $\alpha $ and $u_{22}$ finite, $\alpha $ and $%
u_{22}$ drop out from equations and the bands are decoupled with $\tilde{%
\omega}_{\alpha n}$ and $\tilde{\Delta}_{\alpha n}$ given by $Eqs.(\ref
{omegauni}-\ref{deltauni})$. At $T_{c}$ one has 
\[
\tilde{\omega}_{1n}=\omega _{n}+\Gamma _{1}sign(\omega _{n}) 
\]
\begin{equation}
\tilde{\Delta}_{1n}=\Delta _{1}+\Gamma _{1}(\frac{\tilde{\Delta}_{1n}}{\mid 
\tilde{\omega}_{1n}\mid }+\frac{(1-\sigma )\tilde{\Delta}_{2n}}{\mid \tilde{%
\omega}_{2n}\mid }).  \label{omdel}
\end{equation}
From $Eq.(\ref{omdel})$ it is seen that in the unitarity limit, $\sigma
\rightarrow 1$, the renormalized order parameters are decoupled and $T_{c}$
is unrenormalized. For $\sigma <1$ it can be easily shown that $T_{c}$ is
reduced.

\section{Small anisotropic defect in anisotropic superconductors}

In what follows we consider the effect of a single impurity (small defect)
with small scattering length $a$, which is supposed to be much smaller than
the superconducting coherence length, $\mid a\mid \ll \xi _{0}$. Hence, the
impurity can be considered as a localized perturbation, but with negligible
renormalization of $\hat{\Delta}({\bf p}_{F},{\bf R})$, giving rise to the
quasiclassic equations \cite{Thuneberg}, \cite{Thuneberg2} 
\[
\lbrack (i\omega _{n}+e{\bf v}_{F}\cdot {\bf A(R)})\hat{\tau}_{3}-\hat{\Delta%
}({\bf p}_{F},{\bf R}),\delta \hat{g}({\bf p}_{F},{\bf R},\omega _{n})]+i%
{\bf v}_{F}\nabla _{{\bf R}}\delta \hat{g}({\bf p}_{F},{\bf R},\omega _{n})= 
\]
\begin{equation}
=[\hat{t}({\bf p}_{F},{\bf p}_{F},\omega _{n}),\hat{g}_{imt}({\bf p}_{F},%
{\bf R},\omega _{n})]\delta ({\bf R-R}_{imp}).  \label{eloinhom}
\end{equation}
Here, $\delta \hat{g}({\bf p}_{F},{\bf R},\omega _{n})=\hat{g}({\bf p}_{F},%
{\bf R},\omega _{n})-\hat{g}_{imt}({\bf p}_{F},{\bf R},\omega _{n})$. The
extra term proportional to $\delta ({\bf R-R}_{imp})$ describes a jump in $%
\hat{g}({\bf p}_{F},{\bf R},\omega _{n})$ at the site ${\bf R}_{imp}$ of the
impurity (defect), while the intermediate Green's function $\hat{g}_{imt}(%
{\bf p}_{F},{\bf R},\omega _{n})$ describes the quasiclassic motion in the
absence of impurity (defect) and it is the solution of $Eq.(\ref{eloinhom})$
by putting the right-side to zero. $\hat{g}_{imt}({\bf p}_{F},{\bf R},\omega
_{n})$ is normalized according to $Eq.(\ref{norm})$. The t-matrix is the
solution of $Eq.(\ref{tmatrix})$ where $\hat{g}({\bf p}_{F},\omega _{n})$ is
replaced by $\hat{g}_{imt}({\bf p}_{F},{\bf R=R}_{imp},\omega _{n})$. The
change of the superconducting free-energy in the presence of a single
impurity (defect) is given by \cite{Thuneberg}, \cite{Thuneberg2} 
\begin{equation}
\delta F({\bf R}_{imp})=N(0)T\sum_{n}\int_{0}^{1}d\lambda \int \frac{d^{2}%
\hat{k}_{F}}{4\pi }\int d^{3}RTr[\delta \hat{g}({\bf p}_{F},{\bf R},\omega
_{n})\hat{\Delta}_{b}({\bf p}_{F},{\bf R})],  \label{deltafg}
\end{equation}
where $\hat{\Delta}_{b}({\bf p}_{F},{\bf R})$ and the vector potential ${\bf %
A}_{b}({\bf R})$ are calculated in the absence of the impurity. The Green's
function, $\delta \hat{g}({\bf p}_{F},{\bf R},\omega _{n})$, must be
evaluated for an order parameter $\hat{\Delta}({\bf p}_{F},{\bf R})=\lambda 
\hat{\Delta}_{b}({\bf p}_{F},{\bf R})$.

In the following we study the consequences of anisotropic impurity
scattering for three selected examples of inhomogeneous anisotropic
superconductors.

{\bf 1}. {\bf Bound states due to the anisotropic impurity}

Let us consider the local change of superconductivity in the presence of a
single anisotropic impurity with the potential $v({\bf s,s}^{\prime })$
given by $Eq.(\ref{vsep})$ and analyze the impurity-induced quasiparticle
bound state{\bf \ }and the change in the free-energy $\delta F({\bf R}%
_{imp}) ${\bf . }By assuming that{\bf \ }$2\pi \bar{\sigma}_{i}\ll
E_{F}/\Delta _{0}$ , where $i=0,2$ and $\bar{\sigma}%
_{i}=v_{i}^{2}/(1+v_{i}^{2})$, the t-matrix is given by the same expression
as $Eqs.(\ref{t30}-\ref{t2n})$, but with $\tilde{g}_{2}(n)$ is replaced by $%
g_{2}^{(0)}(n)$. The bound state energy $\omega _{B,anis}<\Delta _{0}$,
which is due to the pair-breaking impurity effects, can be obtained as a
pole of the $t$-matrix which gives 
\begin{equation}
\omega _{B,anis}=\Delta _{0}\sqrt{1-\bar{\sigma}_{pb},}  \label{bound}
\end{equation}
where 
\begin{equation}
\bar{\sigma}_{pb}=\bar{\sigma}_{0}\bar{\sigma}_{2}\frac{(v_{0}-v_{2})^{2}}{%
v_{0}^{2}v_{2}^{2}}.  \label{sigpb}
\end{equation}
In the unitarity limit for both channels, i.e. $v_{0}\gg 1$, $v_{2}\gg 1$
but $v_{2}/v_{0}$ finite, one has $\omega _{B,anis}\rightarrow \Delta _{0}$
contrary to the unitarity limit for the s-wave scattering ($v_{0}\gg 1$, $%
v_{2}=0$) where $\omega _{B,iso}\rightarrow 0$. However, the zero-energy
bound state $\omega _{B,anis}\rightarrow 0$ appears when $v_{0}v_{2}=-1$,
i.e. if one channel is in the unitarity limit the other one must be in the
Born limit.

Due to the bound state there is a change (increase) of the free-energy $%
\delta F({\bf R}_{imp})\equiv \delta F_{imp}$. By solving $Eq.(\ref{eloinhom}%
)$ with $\hat{g}_{imt}({\bf p}_{F},{\bf R},\omega _{n})$ given by $Eq.(\ref
{g0})$ and $\hat{t}$ given by $Eqs.\ref{t3}-\ref{t2n})$ one gets $\delta
F_{imp}$ from $Eq.(\ref{deltafg})$ 
\[
\delta F_{imp}=T\sum_{n}\int_{0}^{1}d\lambda \bar{\sigma}_{pb}\frac{\lambda
\Delta _{0}^{2}\omega _{n}^{2}}{[\omega _{n}^{2}+\lambda ^{2}\Delta
_{0}^{2}][\omega _{n}^{2}+\omega _{B,anis}^{2}]}= 
\]
\begin{equation}
=2T\ln \frac{\cosh (\Delta _{0}/2T)}{\cosh [(1-\sigma _{pb})^{1/2}\Delta
_{0}/2T]},  \label{deltafimp}
\end{equation}
where $\bar{\sigma}_{pb}$ is given in $Eq.(\ref{sigpb})$. It is seen that
there is a loss in the condensation energy,$\delta F({\bf R}_{imp})>0$,
which is related to the pair-breaking effect of impurity. For $v_{0}=v_{2}$
such an impurity does not affect superconductivity and $\delta F({\bf R}%
_{imp})=0$.

The obtained results tell us that in for angle-dependent impurity scattering
even a strong impurity potential may have very weak effect on $T_{c}$, the
bound state, and the free-energy of anisotropic and unconventional pairing.
In that case the anisotropic pairing is robust in the presence of impurities.

{\bf 2. Pinning of single-vortex by a small anisotropic defect}

Because in $HTS$ oxides strong correlations give rise to strong
momentum-dependent charge scattering processes it is interesting to analyze
the elementary-flux-pinning potential of a small defect by using the
approach of Thuneberg et al., \cite{Thuneberg}, \cite{Thuneberg2}, who
showed that in s-wave superconductors the pinning energy of a small defect ($%
a\ll \xi _{0}$) is dominated by scattering processes at the defect. It is
proportional to the product of the scattering cross section and coherence
length ($\propto a^{2}\xi _{0}$), instead of (naively believed) $a^{3}$. The
case of anisotropic pairing with s-wave impurities and near $T_{c}$ was
recently studied in \cite{Friesen}.

In what follows we study the effect of scattering anisotropy on the pinning
energy of a small defect in an anisotropic superconductor at any temperature
below $T_{c}$. We use the model potential given in $Eq.(\ref{vsep})$ and
assume that the vortex is placed at the defect. In order to calculate the
elementary flux-pinning energy one has to solve the quasiclassical equations
for various ballistic trajectories with ${\bf R}$-dependent the vector
potential, ${\bf A(R)}$, and order parameter, 
\begin{equation}
\Delta _{b}({\bf p}_{F},{\bf R})=\mid \Delta ({\bf p}_{F},{\bf R})\mid
e^{i\theta }Y(\theta )  \label{deltaR}
\end{equation}
In the gauge where $\theta $ is the angle with respect to the X-axis then $%
{\bf A(R)}$ has no radial component. The solution of $Eq.(\ref{eloinhom})$
requires for a realistic vortex numerical calculations. For a qualitative
discussion we will adopt a simplified vortex model \cite{Thuneberg}, \cite
{Thuneberg2} which neglects the suppression of the order parameter in the
vortex core and sets $\mid \Delta _{b}({\bf p}_{F},{\bf R})\mid =\Delta _{0}(%
{\bf p}_{F})$, i.e. independent of ${\bf R}$. Hence, the order parameter
along a trajectory passing through the vortex center has constant magnitude
but its phase changes abruptly by $\pi $ when going through the vortex core.
This ''zero-core model'' gives the right order of magnitude of the pinning
energy, $\delta F_{pin}({\bf R}_{imp})$, when compared with the numerical
calculations \cite{Thuneberg2}. In order to calculate $\delta F_{pin}({\bf R}%
_{imp})$ two quantities are needed: $\hat{g}_{imt}({\bf p}_{F},{\bf R=R}%
_{imp},\omega _{n})\equiv \hat{g}_{v}({\bf p}_{F},{\bf R=R}_{imp},\omega
_{n})$ in the presence of the zero-core vortex and the impurity $t$-matrix
calculated with the Green's function, $\hat{g}_{v}({\bf p}_{F},{\bf R=R}%
_{imp},\omega _{n})$, of the zero-core model. The solution is
straightforward \cite{Thuneberg2} and gives 
\begin{equation}
\hat{g}_{v}({\bf p}_{F},{\bf R=R}_{imp},\omega _{n})=\frac{1}{\omega _{n}}%
[(-\Delta _{2}\hat{\tau}_{1}+\Delta _{1}\hat{\tau}_{2})Y(\theta )+(-i\alpha
_{n})\hat{\tau}_{3}],  \label{gv}
\end{equation}
and 
\begin{equation}
\hat{t}({\bf p}_{F},{\bf p}_{F},\omega _{n})=t_{3}\hat{\tau}_{3}=-i\gamma
_{u}\alpha _{n}\omega _{n}[\frac{\bar{\sigma}_{0}}{\omega _{n}^{2}+\bar{%
\sigma}_{0}\Delta _{0}^{2}}+\frac{\bar{\sigma}_{2}}{\omega _{n}^{2}+\bar{%
\sigma}_{2}\Delta _{0}^{2}}]\hat{\tau}_{3}.  \label{tv}
\end{equation}
Here, $\alpha _{n}=\sqrt{\omega _{n}^{2}+\Delta _{0}^{2}}$. $Eq.(\ref
{eloinhom})$ can be solved by the Fourier (or Laplace) transform which gives
the expression for the pinning free-energy 
\begin{equation}
\delta F_{pin}=\delta F_{pin}^{(stiff)}(\bar{\sigma}_{0},\bar{\sigma}%
_{2})+\delta F_{pin}^{(pb)}(\sigma _{pb})  \label{deltafpin}
\end{equation}
\begin{equation}
\delta F_{pin}^{(stiff)}=-2T\ln \{\cosh \frac{\sqrt{\bar{\sigma}_{0}}\Delta
_{0}}{2T}\cdot \cosh \frac{\sqrt{\bar{\sigma}_{2}}\Delta _{0}}{2T}\},
\label{stif}
\end{equation}
\begin{equation}
\delta F_{pin}^{(imp)}=-2T\ln \frac{\cosh (\Delta _{0}/2T)}{\cosh [(1-\bar{%
\sigma}_{pb})^{1/2}\Delta _{0}/2T]}.  \label{pinimp}
\end{equation}
$Eqs.(\ref{deltafpin}-\ref{pinimp})$ imply that, $\delta F_{pin}<0$, and the
vortex is attracted (pinned) by the defect. A comparison of $Eq.(\ref
{deltafpin})$ with the corresponding results for $s-wave$ superconductors
with an s-wave scattering potential shows, that in the former case two
additional terms are present. The first one, depending on $\bar{\sigma}_{2}$%
, appears also in s-wave superconductors with anisotropic scattering
accounted for. In fact $\delta F_{pin}^{(stiff)}$ describes the reduction of
the superconducting stiffness in the presence of impurities. For instance
near $T_{c}$ $Eq.(\ref{stif})$ gives 
\begin{equation}
\delta F_{pin}^{(stiff)}=-(\bar{\sigma}_{0}+\bar{\sigma}_{2})\cdot \frac{%
\Delta _{0}^{2}(T)}{4T_{c}}\approx -7.6\frac{\bar{\sigma}_{0}+\bar{\sigma}%
_{2}}{v_{F}^{2}}\xi _{0}^{2}\cdot T_{c}\Delta _{0}^{2}(T),  \label{stiff}
\end{equation}
and $\delta F_{pin}^{(stiff)}$ is proportional to the total scattering
amplitude $\bar{\sigma}_{0}+\bar{\sigma}_{2}$. For vortex far away from the
impurity there is loss in the condensation energy $\delta F_{pin}^{(pb)}(%
\bar{\sigma}_{pb})$ due to pair-breaking effect of the impurity, i.e. $%
\delta F_{pin}^{(pb)}(\bar{\sigma}_{pb})=-\delta F_{imp}(\bar{\sigma}_{pb})$
where $\delta F_{imp}(\bar{\sigma}_{pb})$ is given by $Eq.(\ref{deltafimp})$%
. Therefore this part enters in $Eq.(\ref{deltafpin})$ with the negative
sign, thus increasing the pinning energy when vortex is sitting on the
defect and stabilizing it additionally. Near $T_{c}$ one has 
\begin{equation}
\delta F_{pin}^{(pb)}(\sigma _{eff})=-\bar{\sigma}_{pb}\cdot \frac{\Delta
_{0}^{2}(T)}{4T_{c}},  \label{cond}
\end{equation}
For the s-wave scattering only, $v_{2}=0$, one has $\bar{\sigma}_{2}=0$, $%
\bar{\sigma}_{pb}=\bar{\sigma}_{0}$, and the pair-breaking effect is maximal
while the condensation energy is gained maximally for vortex sitting on the
defect. However, for, $v_{2}=v_{0}$ the pair-breaking of impurity is absent $%
\bar{\sigma}_{pb}=0$ and $\delta F_{pin}^{(pb)}=0$, i.e. in this case the
pinning by the small defect is similar to that in s-wave superconductors.

The physical picture of the vortex pinning by small defect given above is
based on the known results based on the microscopic derivation of the
Ginzburg-Landau equations in the presence of impurities. An explanation
based on the quasiclassical approach is given in \cite{Thuneberg}, \cite
{Thuneberg2} and we briefly discuss it in order to develop an intuition for
the case of a double-vortex pinning, which is studied below. Because the
order parameter changes its phase by $\pi $ along the trajectories across
the vortex core it leads to the phase change of $\hat{g}_{v}({\bf p}_{F},%
{\bf R},\omega _{n})(\equiv \hat{g}_{imt}({\bf p}_{F},{\bf R},\omega _{n}))$
on the distance $\xi _{0}$, thus causing a cost in the condensation energy,
i.e. the maximal increase of the free-energy. Note, the function $\hat{g}%
_{v}({\bf p}_{F},{\bf R},\omega _{n})$ describes the quasiclassical motion
of particles (or pairs) along trajectories across the vortex core where the
maximal phase change ($\pi $) occurs. In the presence of defect the motion
of particles is described by the function $\hat{g}({\bf p}_{F},{\bf R}%
,\omega _{n})$ which contains scattering of particles to new directions
where the phase change (mismatch) is less than $\pi $ and it costs less
condensation energy. Therefore the vortex is attracted to the defect because
scattering helps superconductivity to sustain abrupt changes in the order
parameter. The latter explains the contribution $\delta F_{pin}^{(stiff)}$,
while in the anisotropic superconductors, due to pair-breaking effects of
impurities, there is a gain of condensation energy $-\delta F_{imp}(\sigma
_{eff})$ for vortex sitting on the defect.

{\bf 3. Pinning of double-vortex by small defect}

We extend the calculations in \cite{Thuneberg2} to the pinning of
multiply-quantized vortices on small defects. First, a s-wave superconductor
is considered and we put the question - is it possible to pin the
double-flux-vortex ($\Phi =2\Phi _{0}$) by the small defect, which is for
simplicity characterized by the parameter $\bar{\sigma}_{0}$ for s-wave
scattering only? The ''zero-core model'' is assumed again. In that case the
order parameter can be parametrized in the form 
\begin{equation}
\Delta _{b}({\bf p}_{F},{\bf R})=\mid \Delta ({\bf p}_{F},{\bf R})\mid
e^{2i\theta }.  \label{delta2R}
\end{equation}
For particle motion across the double-vortex core the order parameter does
not change phase and in that case the solutions for $\hat{g}_{imt}({\bf p}%
_{F},{\bf R=R}_{imp},\omega _{n})(\equiv \hat{g}_{2v}({\bf p}_{F},{\bf R=R}%
_{imp},\omega _{n}))$ and for $\hat{t}({\bf p}_{F},{\bf p}_{F},\omega _{n})$
are given by 
\begin{equation}
\hat{g}_{2v}({\bf p}_{F},{\bf R=R}_{imp},\omega _{n})=\frac{i}{\alpha _{n}}%
[\Delta _{1}\hat{\tau}_{1}+\Delta _{2}\hat{\tau}_{2}+(-\omega _{n})\hat{\tau}%
_{3}]  \label{g2v}
\end{equation}
and 
\begin{equation}
\hat{t}({\bf p}_{F},{\bf p}_{F},\omega _{n})=t_{3}\hat{\tau}_{3}=-i\omega
_{n}\alpha _{n}\gamma _{u}\frac{\bar{\sigma}_{0}}{\omega _{n}^{2}+\tilde{%
\sigma}_{0}\Delta _{0}^{2}}\hat{\tau}_{3},  \label{t2v}
\end{equation}
where $\tilde{\sigma}_{0}=\bar{\sigma}_{0}/v_{0}^{2}$. Then by solving $Eq.(%
\ref{eloinhom})$ and by using $Eq.(\ref{deltafg})$ and this solutions one
obtains the pinning energy of the double-vortex within the zero-core model 
\begin{equation}
\delta F_{2v,pin}=2T\sum_{n}\int_{0}^{1}d\lambda \frac{\lambda \Delta
_{0}^{2}\omega _{n}^{2}\bar{\sigma}_{0}}{\alpha _{n}^{2}[\omega _{n}^{2}+%
\tilde{\sigma}_{0}\lambda ^{2}\Delta _{0}^{2}]}>0.  \label{f2vpi}
\end{equation}
The main conclusion coming out from $Eq.(\ref{f2vpi})$ is that because $%
\delta F_{pin}>0$ the double-vortex in s-wave superconductors is repelled
from the defect - i.e. the zero-core double-vortex can not be pinned.
Contrary to the single-vortex, where the defect scatters particles to new
directions where the phase change is smaller, in the case of double-vortex
the particles are scattered to directions where the phase change is larger.
However, it might be that the above obtained results are an artefact of the
''zero-core model'', where there is no suppression of the superconducting
order due to the vortex core, and numerical calculations are required for a
realistic double-vortex structure \cite{Endres}.

In the case of an unconventional pairing, like that in Section $II.1$, the
order parameter is given by $\Delta _{b}({\bf p}_{F},{\bf R})=\mid \Delta (%
{\bf p}_{F},{\bf R})\mid \exp (2i\theta )Y(\theta )$ and the t-matrix
contains also terms $t_{1},t_{2}\neq 0$ leading to a decrease of the jump $[%
\hat{t}({\bf p}_{F},{\bf p}_{F},\omega _{n}),\hat{g}_{imt}({\bf p}_{F},{\bf R%
},\omega _{n})]$ in $Eq.(\ref{eloinhom})$. In this case the pinning energy
contains the additional term (gain in energy) due to the pair-breaking
effect of the impurity, $-\delta F_{imp}$, i.e. $\delta F_{pin}=\delta
F_{2v,pin}-\delta F_{imp}$. Since $\delta F_{2v,pin}$ is less positive
(repulsive) than for a s-wave superconductor in $Eq.(\ref{f2vpi})$, and
because $-\delta F_{imp}<0$ one can happen that $\delta F_{pin}<0$\ and even
the zero-core double-vortex can be pinned by the defect. A realistic
calculation of $\delta F_{pin}$ for anisotropic superconductors with
anisotropic scattering of single and double-vortex will be discussed
elsewhere \cite{Endres}.

In conclusion the anisotropic impurity scattering gives rise to new
qualitative effects in unconventional and anisotropic superconductors, where
for instance it ''screens'' the strength of the scattering in some
quantities (like $T_{c}$ - robustness of pairing, bound states, pinning,
etc.) even in the unitarity limit. It seems that this situation is partly
realized in $HTS$ oxides where the d-wave pairing is robust in the presence
of even very strong impurity scattering. In two-band models nonmagnetic
impurities do not affect thermodynamic properties of s-wave superconductors
in the unitarity limit for the interband scattering, contrary to the Born
limit, i.e. in this case the Anderson theorem is restored in the unitarity
limit.

{\bf Acknowledgments}

M. L. K. thanks Dierk Rainer for valuable discussions, suggestions and for
careful reading and correcting the manuscript. M. L. K. acknowledges
gratefully the support of the Deutsche Forschungsgemeinschaft through the
Forschergruppe ''Transportph\"{a}nomene in Supraleitern und Suprafluiden''.
O.V. D. thanks Nils Schopohl for valuable discussions and support.

\end{document}